\documentstyle[pra,aps]{revtex}

\newcommand{\beq}{\begin{equation}}
\newcommand{\eeq}{\end{equation}}
\newcommand{\bea}{\begin{eqnarray}}
\newcommand{\eea}{\end{eqnarray}}

\draft

\begin{document}
\title{Line narrowing via cavity-induced quantum interference in a $\Xi $-type atom}
\author{Peng Zhou\thanks{Electronic address: peng.zhou@physics.gatech.edu}}
\address{School of Physics, Georgia Institute of Technology, Atlanta, GA 30332-0430, USA.\\
Department of Physics, Hunan Normal University, Changsha 410081,
China.}
\date{September 8, 1999}
\maketitle

\begin{abstract}
We show that cavity-induced interference may result in spectral line
narrowing in the absorption spectrum of a $\Xi$-type atom coupled to a
single-mode, frequency-tunable cavity field with a pre-selected polarization
at finite temperature.
\end{abstract}


Within recent years, quantum interference among different
transition pathways of atoms has become a very important tool in
manipulating radiative properties of atoms \cite{arim96}. Many
interesting and counterintuitive effects, such as
electromagnetically-induced transparency \cite{EIT}, enhancement
of the index of refraction without absorption \cite{index}, lasing
without population inversion \cite{harris}, fluorescence quenching
\cite{cardimona,zhou1}, quantum beats \cite{beats}, are attributed
to the interference. Very recently, we have shown that quantum
interference can also result in spectral line narrowing
\cite{zhou1,nar} and is potentially of interest as a spectroscopic
tool, as well.

Quantum interference effects can be generated by coherent laser fields \cite
{EIT,index,nar,scully}. For example, for three-level atomic systems (in $V$,
$\Lambda $ and $\Xi $ configurations) excited by two laser fields: one being
a strong pump field to drive two levels (say $|1\rangle $ and $|2\rangle $)
and the other being a weak probe field at different frequency to probe the
levels $|0\rangle $ and $|1\rangle $ or $|2\rangle $, the strong coherent
field can drive the levels $|1\rangle $ and $|2\rangle $ into superpositions
of these states, so that different atomic transitions are correlated. For
such systems, the cross-transition terms (quantum interference) are evident
in the atomic dressed picture \cite{EIT,index,nar,scully}. The interference
may also arise as two transition pathways are created due to atomic emission
into a common vacuum of the infinite electromagnetic modes \cite
{harris,cardimona,zhou1,jan}. The latter requires that the dipoles moments
of the transitions involved are non-orthogonal. The maximal effect of
quantum interference occurs with parallel moments. From the experimental
point view, however, it is difficult to find isolated atomic systems which
have parallel moments \cite{harris,cardimona,jan,berman,agarwal}.

An alternative scheme for engineering quantum interference (equivalently two
parallel or anti-parallel dipole transition moments) has been proposed by
coupling a V-type atom to a cavity mode with low quality factor $Q$ \cite
{agarwal}. Here we extend the study to a $\Xi $-type atom coupled to a
frequency-tunable, single-mode cavity field with a pre-selected polarization
at finite temperature. We show that maximal quantum interference can be
achieved in such a system, and the cavity-induced interference may lead to
linewidth narrowing in the probe absorption spectrum.

For a $\Xi $-configuration three-level atom having the levels $|0\rangle $, $%
|1\rangle $ and $|2\rangle $ with level energies $E_{0}<E_{1}<E_{2}$, we
assume that the levels $|0\rangle \leftrightarrow |1\rangle $, and $%
|1\rangle \leftrightarrow $ $|2\rangle $ are coupled by the single-mode
cavity field, respectively. Direct transitions between the ground level $%
|0\rangle $ and the upper excited level $|2\rangle $ are dipole forbidden.
The master equation for the total density matrix operator $\rho _{T}$ in the
frame rotating with the average atomic transition frequency $\omega
_{0}=(E_{2}-E_{0})/2$ takes the form,

\begin{equation}
\dot{\rho}_{T}=-i\left[ H_{A}+H_{C}+H_{I},\,\rho _{T}\right] +{\cal L}\rho
_{T},  \label{master}
\end{equation}
where
\begin{eqnarray}
H_{A} &=&\Delta A_{11}, \\
H_{C} &=&\delta \,a^{\dagger }a, \\
H_{I} &=&i\left( g_{01}A_{01}+g_{12}A_{12}\right) a^{\dagger }-h.c., \\
{\cal L}\rho _{T} &=&\kappa (N+1)\left( 2a\rho _{T}a^{\dagger }-a^{\dagger
}a\rho _{T}-\rho _{T}a^{\dagger }a\right)   \nonumber \\
&&+\kappa N\left( 2a^{\dagger }\rho _{T}a-aa^{\dagger }\rho _{T}-\rho
_{T}aa^{\dagger }\right) ,
\end{eqnarray}
where $H_{C}$, $H_{A}$ and $H_{I}$ are the unperturbed cavity, the
unperturbed atom and the cavity-atom interaction Hamiltonians respectively,
while ${\cal L}\rho _{T}$ describes damping of the cavity field by the
continuum electromagnetic modes at finite temperature, characterized by the
decay constant $\kappa $ and the mean number of thermal photons N; $a$ and $%
a^{\dag }$ are the photon annihilation and creation operators of the cavity
mode, and $A_{ij}=|i\rangle \langle j|$ is the atomic population (the dipole
transition) operator for $i=j$ $(i\neq j)$; $\delta =\omega _{C}-\omega _{0}$
is the cavity detuning from the average atomic transition frequency, $\Delta
=E_{1}-\omega _{0}$, and $g_{ij}={\bf e}_{\lambda }\cdot {\bf d}_{ij}\sqrt{%
\hbar \omega _{C}/2\epsilon _{0}V}$ is the atom-cavity coupling constant
with ${\bf d}_{ij},$ the dipole moment of the atomic transition from $%
|j\rangle $ to $|i\rangle ,$ ${\bf e}_{\lambda }$, the polarization of the
cavity mode, and $V,$ the volume of the system. In the remainder of this
work we assume that the polarization of the cavity field is {\em pre-selected%
}, i.e., the polarization index $\lambda $ is fixed to one of two possible
directions.

In this paper we are interested in the bad cavity limit: $\kappa \gg g_{ij} $%
, that is the atom-cavity coupling is weak, and the cavity has a low $Q$ so
that the cavity field decay dominates. The cavity field response to the
continuum modes is much faster than that produced by its interaction with
the atom, so that the atom always experiences the cavity mode in the state
induced by the thermal reservoir. Thus one can adiabatically eliminate the
cavity-mode variables, giving rise to a master equation for the atomic
variables only \cite{detail}, which are of the form

\begin{eqnarray}
\dot{\rho} &=&-i\left[ H_{A},\;\rho \right]  \nonumber \\
&&+\{F(-\Delta )(N+1)\left[ |g_{01}|^{2}\left( A_{01}\rho A_{10}-A_{11}\rho
\right) +g_{01}g_{12}^{*}A_{01}\rho A_{21}\right]  \nonumber \\
&&+F(\Delta )(N+1)\left[ |g_{12}|^{2}\left( A_{12}\rho A_{21}-A_{22}\rho
\right) +g_{01}^{*}g_{12}A_{12}\rho A_{10}\right]  \nonumber \\
&&+F(-\Delta )N\left[ |g_{01}|^{2}\left( A_{10}\rho A_{01}-\rho
A_{00}\right) +g_{01}g_{12}^{*}A_{21}\rho A_{01}\right]  \nonumber \\
&&+F(\Delta )N\left[ |g_{12}|^{2}\left( A_{21}\rho A_{12}-\rho A_{11}\right)
+g_{01}^{*}g_{12}A_{10}\rho A_{12}\right]  \nonumber \\
&&+h.c.\},  \label{master1}
\end{eqnarray}
where $F(\pm \Delta )=[\kappa +i(\delta \pm \Delta )]^{-1}$

Obviously, the equation (\ref{master1}) describes the cavity-induced atomic
decay into the cavity mode. The real part of $F(\pm \Delta )|g_{ij}|^{2}$
represents the cavity-induced decay rate of the atomic level $|j\rangle $ to
the lower level $|i\rangle \,$, while the imaginary part is associated with
the frequency shift of the atomic level resulting from the interaction with
the thermal field in the detuned cavity. The other terms, $F(\pm \Delta
)g_{01}g_{12}^{*}$ and $F(\pm \Delta )g_{01}^{*}g_{12}\,$, however,
represent the cavity-induced correlated transitions of the atom, in which
the atomic transition $|1\rangle \rightarrow |0\rangle $ induces the other
transition $|1\rangle \rightarrow |2\rangle $, and vice versa. It is these
correlated transitions that give rise to quantum interference.

The effect of quantum interference is very sensitive to the orientations of
the atomic dipoles and the polarization of the cavity mode. For instance, if
the cavity-field polarization is not pre-selected, as in free space, one
must replace $g_{ij}g_{kl}^{*}$ by the sum over the two possible
polarization directions, giving $\Sigma _{\lambda }g_{ij}g_{kl}^{*}\propto
{\bf d}_{ij}\cdot {\bf d}_{kl}^{*}$ \cite{agarwal}. Therefore, only
non-orthogonal dipole transitions lead to nonzero contributions, and the
maximal interference effect occurs with the two dipoles parallel. As pointed
out in Refs. \cite{harris,cardimona,berman,agarwal} however, it is
questionable whether there is a isolated atomic system with parallel
dipoles. Otherwise, if the polarization of the cavity mode is fixed, say $%
{\bf e}_{\lambda }={\bf e}_{x}$, the polarization direction along the $x$%
-quantization axis, then $g_{ij}g_{kl}^{*}\propto \left( {\bf d}_{ij}\right)
_{x}\left( {\bf d}_{kl}^{*}\right) _{x}$, which is nonvanishing, regardless
of the orientation of the atomic dipole matrix elements.

Now we investigate the effects of quantum interference on the steady-state
absorption spectrum of such a system, which is defined as

\begin{equation}
A(\omega )=\Re e\int_{0}^{\infty }\lim_{t\rightarrow \infty }\left\langle
\left[ P(t+\tau ),\;P^{\dag }(t)\right] \right\rangle e^{i\omega \tau }d\tau
,  \label{spec}
\end{equation}
where $\omega =\omega _{p}-\omega _{0}$, and $\omega _{p}$ is the frequency
of the probe field and $P(t)=d_{1}A_{01}+d_{2}A_{12}$ is the component of
the atomic polarization operator in the direction of the probe field
polarization vector ${\bf e}_{p}$, with $d_{1}={\bf e}_{p}\cdot {\bf d}_{01}$
and $d_{2}={\bf e}_{p}\cdot {\bf d}_{12}$. Obviously, we may probe atomic
absorption between different transition levels by selecting the polarization
of the probe beam \cite{jan}. For example, if the polarization direction $%
{\bf e}_{p}$ is perpendicular to the dipole moment ${\bf d}_{12}$ of atomic
transitions between the levels $|1\rangle $ and $|2\rangle $, then $d_{2}=0$%
, the spectrum (\ref{spec}) hence describes the atomic absorption between
the ground level $|0\rangle $ and the intermediate level $|1\rangle $, while
it represents the probe absorption from the intermediate level $|1\rangle $
to the upper level $|2\rangle $, if ${\bf e}_{p}\perp {\bf d}_{01}$,
otherwise, eq. (\ref{spec}) is the stepwise two-photon absorption spectrum
between the ground and upper levels.

The spectrum (\ref{spec}) can be evaluated from the cavity-modified Bloch
equations

\begin{eqnarray}
\dot{\rho}_{11} &=&-2\gamma _{1}\left[ (N+1)\rho _{11}-N\rho _{00}\right]
-2\gamma _{2}\left[ N\rho _{11}-(N+1)\rho _{22}\right] ,  \nonumber \\
\dot{\rho}_{22} &=&-2\gamma _{2}\left[ (N+1)\rho _{22}-N\rho _{11}\right] ,
\nonumber \\
\dot{\rho}_{20} &=&-\left[ \gamma _{1}N+\gamma _{2}(N+1)+i\omega
_{sh}\right] \rho _{20},  \nonumber \\
\dot{\rho}_{10} &=&-\left[ F(-\Delta )|g_{01}|^{2}(2N+1)+F^{*}(\Delta
)|g_{12}|^{2}N+i\Delta \right] \rho _{10}  \nonumber \\
&&+\left[ F(\Delta )+F^{*}(-\Delta )\right] g_{01}^{*}g_{12}(N+1)\rho _{21},
\nonumber \\
\dot{\rho}_{21} &=&-\left[ F^{*}(-\Delta )|g_{01}|^{2}N+F(\Delta
)|g_{12}|^{2}(2N+1)-i\Delta \right] \rho _{21}  \nonumber \\
&&+\left[ F^{*}(\Delta )+F(-\Delta )\right] g_{01}g_{12}^{*}N\rho _{10},
\label{bloch}
\end{eqnarray}
where $\omega _{sh}=\Im m\left[ F(-\Delta )\right] N|g_{01}|^{2}+\Im m\left[
F(\Delta )\right] (N+1)|g_{12}|^{2}$ is the cavity-induced frequency shift,
and $\gamma _{1}$ ($\gamma _{2}$) is the cavity-induced decay rate of the
intermediate state $|1\rangle $ to the ground state $|0\rangle $ ($|2\rangle
\rightarrow |1\rangle $),

\begin{equation}
\gamma _{1}=\frac{\kappa |g_{01}|^{2}}{\kappa ^{2}+(\delta -\Delta )^{2}}%
,\;\;\;\gamma _{2}=\frac{\kappa |g_{12}|^{2}}{\kappa ^{2}+(\delta +\Delta
)^{2}},  \label{rate}
\end{equation}
which vary with the cavity frequency, decay rate and coupling constants. It
is not difficult to find from eq. (\ref{bloch}) that the steady state
populations are, however, independent of these parameters:

\begin{eqnarray}
\rho _{22} &=&\frac{N^{2}}{3N(N+1)+1},  \nonumber \\
\rho _{11} &=&\frac{N(N+1)}{3N(N+1)+1},  \nonumber \\
\rho _{00} &=&\frac{(N+1)^{2}}{3N(N+1)+1},
\end{eqnarray}
and the steady-state coherence between the ground and upper levels is zero, $%
\rho _{20}=0$, which are the same as a $\Xi $-type atom damped by a thermal
reservoir in free space. Nevertheless, the transit atomic coherence $\rho
_{20}(t)$ may oscillate due to the cavity-induced frequency shift.

Figure 1 represents the probe absorption spectrum between the ground state $%
|0\rangle $ and the intermediate state $|1\rangle $, for $\Delta =\delta =0$%
, $g_{01}=g_{12}=g=10$, $\kappa =100$, and various numbers of thermal
photon. It is evident that when the photon number is very small, see $N=0.01$
in Fig. 1(a) for instance, both the spectra in the presence and absence of
the cavity induced interference are virtually same. Otherwise, the spectral
line is narrowed in the presence of the interference, as shown in Figs.
1(b)-1(d), where $N=0.1,\,1$, and $2$, respectively. The larger the number, the
more profound the line narrowing. One can find from eq. (\ref{bloch}) that
in the case, no frequency shift occurs, and the absorption spectrum consists
of two Lorentzians centred at $\omega _{p}=\omega _{0}$ with the spectral
linewidths $\Gamma _{\pm }=2\,[(3N+1)\pm 2\eta \sqrt{N(N+1)}]g^{2}/\kappa $
. In the absence of the cavity-induced quantum interference ($\eta =0$), the
two Lorentzians are same. Otherwise, $\eta =1$, the Lorentzians have
different linewidths.. Noting that heights of the Lorentzians are
proportional to $(1/\Gamma _{\pm })$, the Lorentzian with linewidth $%
\Gamma _{-}$ dominates. Therefore, the absorption peck is narrower and
higher in the presence of the cavity-induced quantum interference.

Figure 2 shows the variations of the spectral line narrowing with the cavity
detuning for $N=1$ and $\Delta =0$. The spectrum is sharper in large cavity
detuning, for example, $\delta =100$ in the frame 2(d). This is because the
cavity-induced decay rates, as shown in eq. (\ref{rate}), are suppressed
when the cavity is far off resonant with the atomic transition. In addition,
the absorption peak is displaced a bit from the atomic transition frequency
involved $\omega _{10}=\omega _{0}$ for large detuning, owing to the cavity
induced frequency shifts.

Figure 3 exhibits that the effect of the cavity induced interference on the
spectral linewidth narrowing is significant only when the $\Xi $-type atom
has an equally-spaced or near equally-spaced level structure, as displayed
in Figs. 3(a) with $\Delta =0$, and 3(b) with $\Delta =1$. Otherwise, the
effect is negligible small, which is demonstrated in the frames 3(c) and
3(d) for $\Delta =5$ and $10$, respectively.

It should be pointed out that the features of the cavity interference
assisted line narrowing in the probe spectrum between the intermediate and
upper states, and in the stepwise two-photon absorption spectrum between the
ground and upper states are similar to those in the atomic absorption
spectrum from the ground state to the intermediate state, as exhibited in
the above figures.

In summary, we have shown that cavity-induced quantum interference can lead
to significant linewidth narrowing in the probe absorption spectra of a $\Xi
$-type atom with equal (or near equal) space level-lying, coupled to a
single-mode, frequency-tunable cavity field, which is damped by a thermal
reservoir, in the bad cavity limit. There are no special restrictions on the
atomic dipole moments for the appearance of quantum interference , as long
as the polarization of the cavity field is pre-selected.

\acknowledgments

This work is supported by the ARO/NSA grant G-41-Z05. I gratefully
acknowledge conversations with Z. Ficek, T. A. B. Kennedy,  S.
Swain, and L. You.

\begin{figure}[tbp]
\caption{Absorption spectrum between the ground state $|0\rangle$ and the
intermediate state $|1\rangle$, for $\Delta =\delta=0$, and $N
=0.01,\,0.1,\,1,\,2$ in (a)--(d), respectively. In all figures, $%
g_{01}=g_{12}=10$ and $\kappa =100$ are set, and the solid curves represent
the spectrum in the presence of the cavity-induced interference, whilst the
dashed curves are the spectrum in the absence of the interference. }
\label{fig1}
\end{figure}

\begin{figure}[tbp]
\caption{Same as FIG. 1, but with $N=1,\,\Delta =0$, and $\delta
=0,\,10,\,50,\,100$ in (a)--(d), respectively. }
\label{fig2}
\end{figure}
\begin{figure}[tbp]
\caption{Same as FIG. 1, but with $N=1,\,\delta =0$, and $\Delta
=0,\,1,\,5,\,10$ in (a)--(d), respectively. }
\label{fig3}
\end{figure}

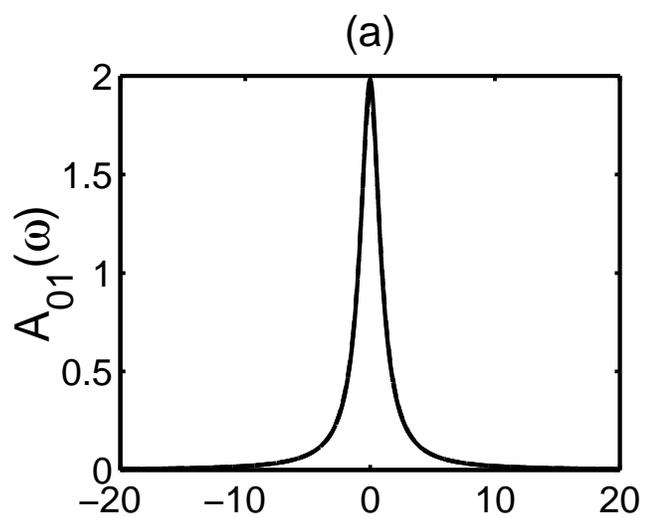
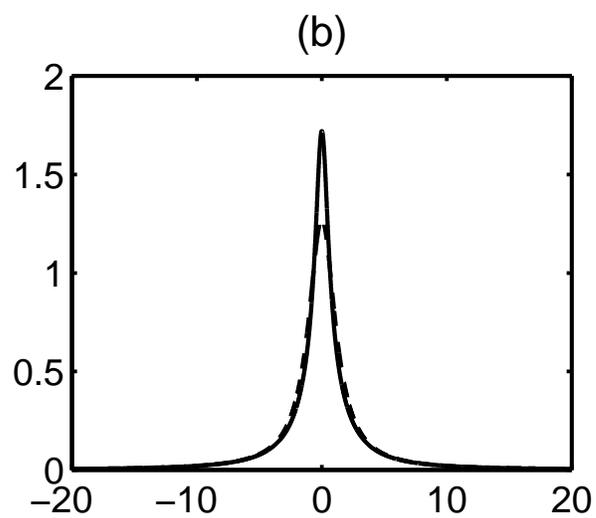
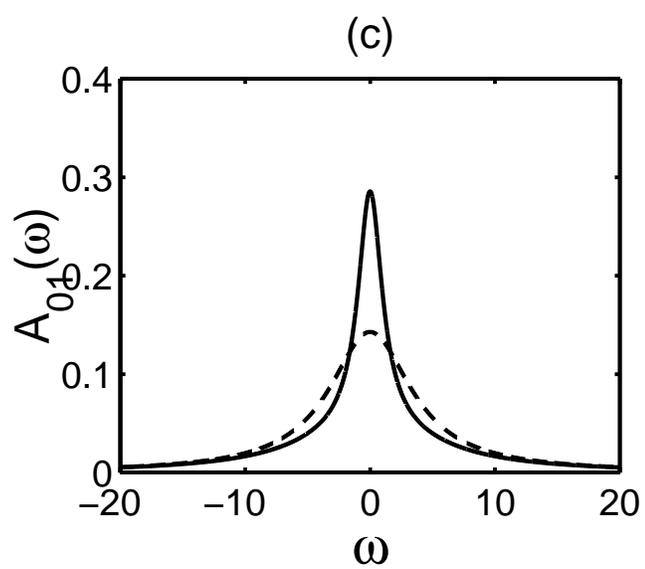
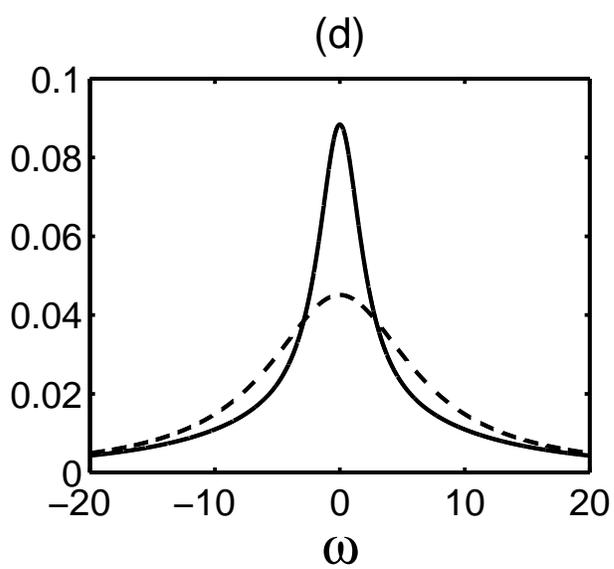
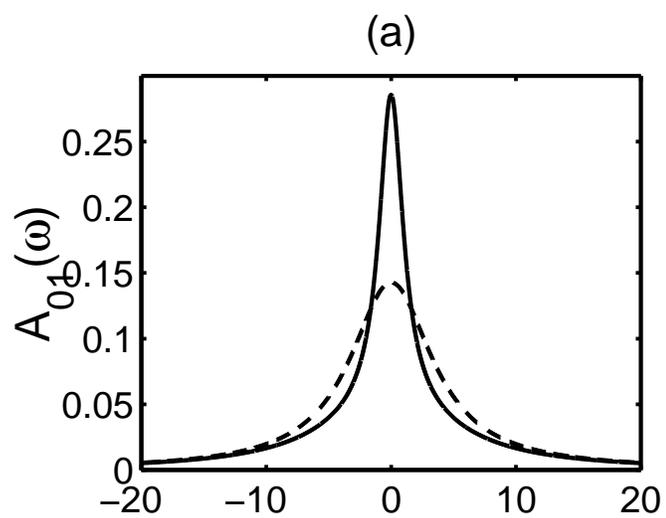
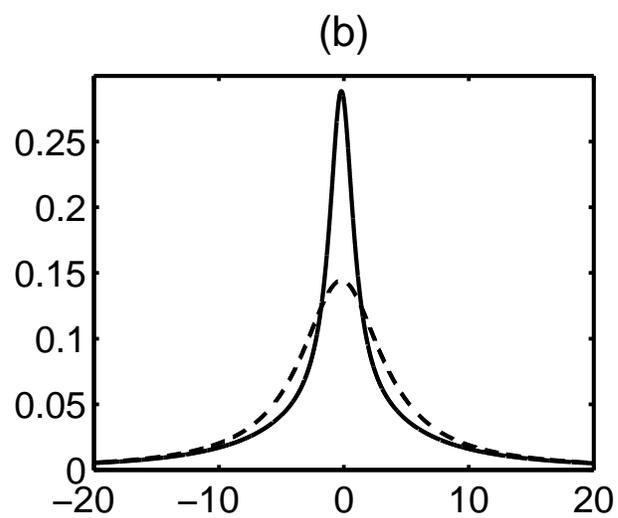
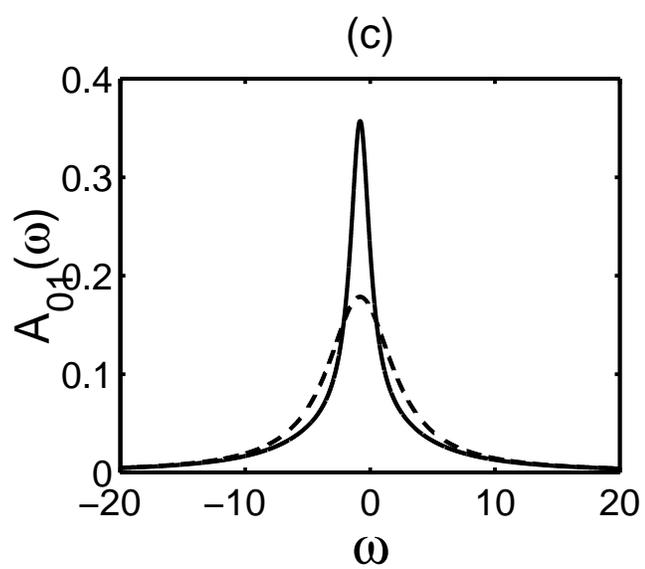
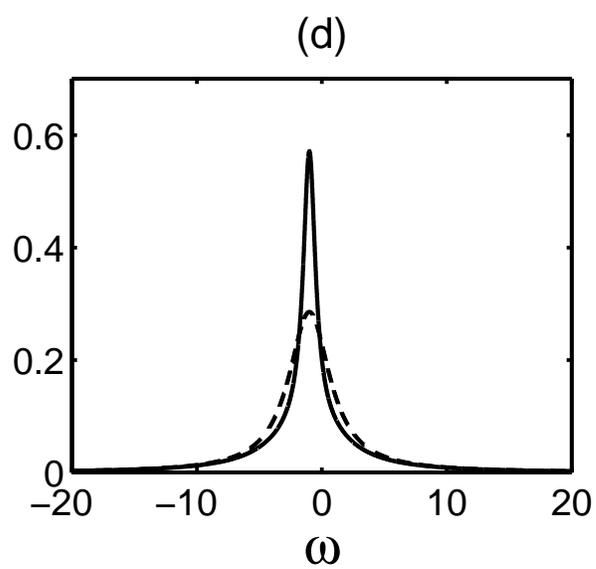
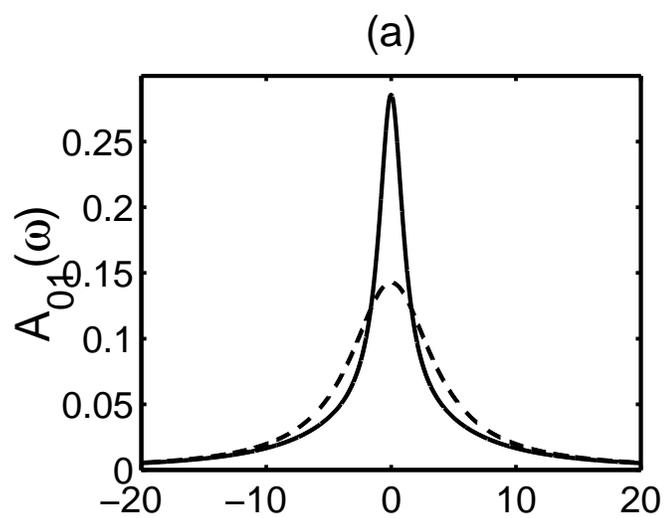
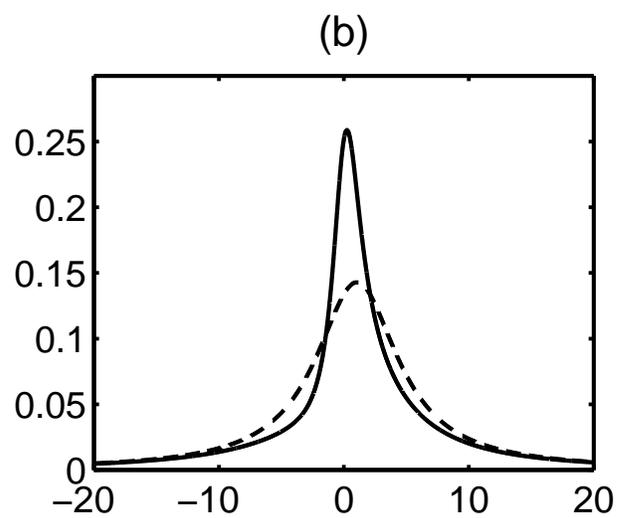
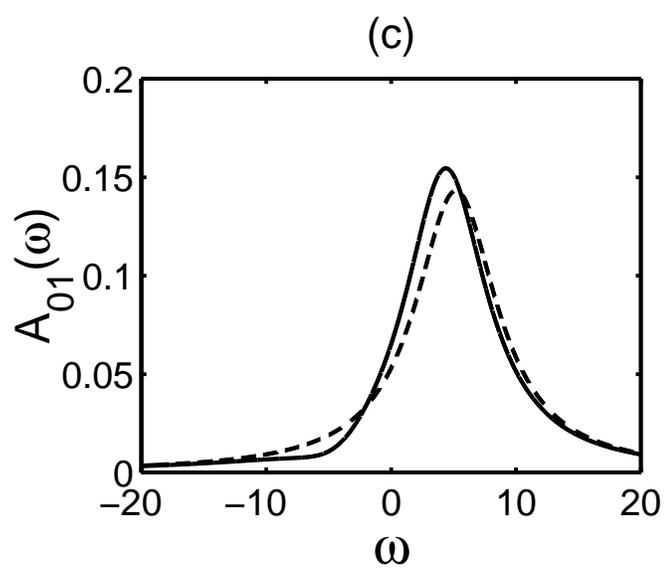
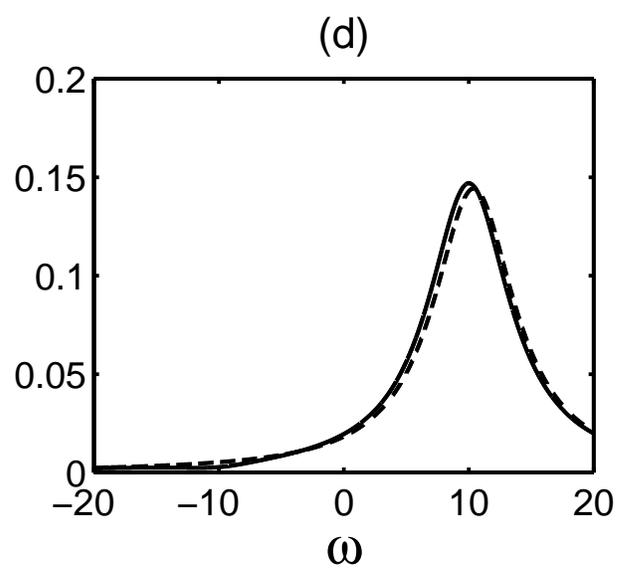
\end{document}